\documentclass[10pt,conference]{IEEEtran}
\usepackage{cite}
\usepackage{amsmath,amssymb,amsfonts}
\usepackage{algorithmic}
\usepackage{graphicx}
\usepackage{textcomp}
\usepackage{xcolor}
\usepackage{listings}
\usepackage{multirow}
\usepackage{tikz}
\def\BibTeX{{\rm B\kern-.05em{\sc i\kern-.025em b}\kern-.08em
    T\kern-.1667em\lower.7ex\hbox{E}\kern-.125emX}}

\begin{document}

\title{Developers Are Victims Too : A Comprehensive Analysis of The VS Code Extension Ecosystem}

\author{
    \IEEEauthorblockN{Shehan Edirimannage\textsuperscript{1*}, Charitha Elvitigala\textsuperscript{1*}, Asitha Kottahachchi Kankanamge Don\textsuperscript{1}, Wathsara Daluwatta\textsuperscript{1},},
    \IEEEauthorblockN{Primal Wijesekara\textsuperscript{2}, Ibrahim Khalil\textsuperscript{1}}
    
    \IEEEauthorblockA{ \textsuperscript{1} \textit{School of Computing Technologies, RMIT University, Melbourne, VIC 3000, Australia}}
    \IEEEauthorblockA{\textsuperscript{2} \textit{International Computer Science Institute \& University of California, Berkeley, CA, 94704, US}}
    \IEEEauthorblockA{\textsuperscript{*} The first two authors contributed equally to this work.}
    \IEEEauthorblockA{\{shehan.edirimannage, charitha.elvitigala, asitha.kottahachchi.kankanamge.don, wathsara.daluwatta\}@student.rmit.edu.au}
    \IEEEauthorblockA{primalw@icsi.berkeley.edu, ibrahim.khalil@rmit.edu.au}
}

\newcommand*\circled[1]{\tikz[baseline=(char.base)]{
            \node[shape=circle,draw,inner sep=1pt] (char) {#1};}}

\newcommand{\added}[1]{\textcolor{red}{#1}}

\definecolor{keys}{HTML}{1a7f37}
\definecolor{values}{HTML}{0969da}

\lstset{
    basicstyle=\ttfamily \footnotesize,
    string=[s]{"}{"},
    stringstyle=\color{keys},
    showstringspaces=false,
    literate=
        {:=}{{\textcolor{red}{$\gets$}}}1
        {"onCommand:me.sayHello"}{{\textcolor{values}{"onCommand:me.sayHello"}}}{23}
        {"chrismarti.regex"}{{\textcolor{values}{"chrismarti.regex"}}}{18}
        {"zobo.php"}{{\textcolor{values}{"zobo.php"}}}{10}
        {11}{{\textcolor{red}{\circled{1}}}}{1}
        {12}{{\textcolor{red}{\circled{2}}}}{1}
        {13}{{\textcolor{red}{\circled{3}}}}{1}
        {14}{{\textcolor{red}{\circled{4}}}}{1}
        {15}{{\textcolor{red}{\circled{5}}}}{1}
        {\{}{{{\color{black}{\{}}}}{1}
        {\}}{{{\color{black}{\}}}}}{1}
        {[}{{{\color{black}{[}}}}{1}
        {]}{{{\color{black}{]}}}}{1},
    comment=[l]{:},
    commentstyle=\color{values},
    captionpos=b
}

\newenvironment{packed_enum}{
\begin{enumerate}
  \setlength{\itemsep}{1pt}
  \setlength{\parskip}{0pt}
  \setlength{\parsep}{0pt}
}{\end{enumerate}}

\newenvironment{packed_item}{
\begin{itemize}
  \setlength{\itemsep}{1pt}
  \setlength{\parskip}{0pt}
  \setlength{\parsep}{0pt}
}{\end{itemize}}

\maketitle

\begin{abstract}

With the wave of high-profile supply chain attacks targeting development and client organizations, supply chain security has recently become a focal point. As a result, there is an elevated discussion on securing the development environment and increasing the transparency of the third-party code that runs in software products to minimize any negative impact from third-party code in a software product. However, the literature on secure software development lacks insight into how the third-party development tools used by every developer affect the security posture of the developer, the development organization, and, eventually, the end product. To that end, we have analyzed 52,880 third-party VS Code extensions to understand their threat to the developer, the code, and the development organizations. We found that ~5.6\% of the analyzed extensions have suspicious behavior, jeopardizing the integrity of the development environment and potentially leaking sensitive information on the developer's product. We also found that the VS Code hosting the third-party extensions lacks practical security controls and lets untrusted third-party code run unchecked and with questionable capabilities. We offer recommendations on possible avenues for fixing some of the issues uncovered during the analysis.

\end{abstract}


\section{Introduction}

Accepting the \textit{Turing Award} in 1984, Thompson projected that reliance on third-party software components involves a significant amount of trust~\cite {thompson1984reflections}. Despite this warning and subsequent research and incidents stemming from supply chain issues~\cite{solarwind, rapid7, dragonfy, junpier, peisert2021perspectives, vancouver}, developers' and organizations' reliance on third-party components continues to increase. At the same time, these components create security risks through undocumented functionality and settings, bugs, or malicious code.

Supply chain attacks have emerged as a pervasive and ubiquitous threat vector in recent years~\cite{spinellis2003reflections, DBLP:journals/ieeesp/BratusDLPSS14, levy2003poisoning}. The main reason why these types of attacks can be catastrophic is due to the nature of their coverage: numerous systems and software products may be vulnerable, often without users', administrators', or organizations' knowledge, as most software products do not enumerate their constituent third-party components. Mounting an attack on several hundred organizations would otherwise be daunting; however, if the attacker can compromise one of the tools or systems used by all those organizations, then the economics suddenly shift heavily to the attacker. This has become quite apparent in the wake of several high-profile attacks~\cite{solarwind, peisert2021perspectives, log4j}. The sheer scale of these attacks is unprecedented. However, the security community has been aware of these risks for almost 40 years.

Standards, like the Software Bill of Materials (SBoM), increase transparency and awareness of the third-party code that may be executed in an organizational environment.
A few of the current literature on supply chain security includes securing CI/CD~\cite{sec22_koishybayev_GithubCI, CICD_Malware} or securing update channels~\cite{supply_chain_security_30_industry} focusing on securing the product even before it ships out of the development organization. Security literature has looked into how developers write secure code~\cite{fischer2017stack} or how they look for security advice~\cite{acar2016you}, which is essential in securing the whole ecosystem. However, the recent high-profile attacks on a development organization emphasize the importance of securing the roots of the supply chain: the developer~\cite{Satter2023MicrosoftHack}.  

While SBoM helps to articulate the composition of the software code consisting of third-party software libraries, there has been an oversight on one aspect of the development activity that involves third-party code: developer tools. Developers use tools for a wide range of activities that have a direct impact on the final software code. There is already documented evidence~\cite{goldman2023trust, checkpoint2023vscode} of malicious developer tools that could jeopardize the integrity of the development environment, which will lead to catastrophic events such as recent attacks~\cite{Satter2023MicrosoftHack}. However, security and privacy literature lacks systematic knowledge of how third-party developer tools behave in the wild and their impact on the security posture of the developer, the organization, and the end product.

To that end, we analyzed 52,880 third-party VS Code extensions using static and dynamic analysis methods. We used static analysis to examine the code and narrow the extension list with suspicious code segments. We then executed those selected extensions using an instrumented VS Code environment, logging all their execution aspects. We also used VirusTotal~\cite{VirusTotal}, Retire.js~\cite{RetireJSRepository} to scan for malicious content and known vulnerabilities. With this holistic analysis approach, we found that 5.6\%~\footnote{https://github.com/vulnerability-reporter/DAV2-ACAnTVSCEE} of our collected extension set has suspicious behavior that could jeopardize the integrity of the development environment and/or leak sensitive information such as code and personally identifiable information.

We contribute the following:
\begin{itemize}
    \item Systematically show that the third-party developer tools pose a serious security threat to the developer, the host computer, and the organization.
    \item VS Code, one of the most popular developer tools, has a lax security architecture that lets third-party extensions run unchecked, resulting in serious security lapses.
    \item To the best of our knowledge, the paper presents the first holistic (52, 880) analysis of developer tools and their security and privacy implications.
\end{itemize}

\section{Visual Studio Code}

Visual Studio Code (VS Code)~\cite{VSCodeEditor} is a free, open-source code editor developed by Microsoft. VS Code is built on the Electron framework~\cite{ElectronJS}, which facilitates the creation of cross-platform desktop applications. Electron leverages the Chromium engine~\cite{Chromium} for rendering web pages within applications. The Stack Overflow 2023 Developer Survey~\cite{StackOverflowSurvey2023} reveals that Visual Studio Code continues to be the preferred IDE among developers. VS Code supports various programming languages, including JavaScript, TypeScript, Python, PHP, C++, and C\#. 
\subsection{Extensions}

A Visual Studio Code (VS Code) extension is a piece of third-party software installed into the VS Code editor to enhance its functionality. The development of VS Code extensions is a collaborative effort involving the community and technology providers. A VS Code extension can be defined as a Node.js application. Each Node.js application contains a \texttt{package.json} file, which holds descriptive and functional metadata about the application. Similarly, a VS Code extension includes a \texttt{package.json} file that contains information specific to the extension~\cite{VSCodeExtensionManifest}. This file holds additional details pertinent to the extension, akin to the Manifest files found in Android apps and Chrome extensions.

A portion of a sample package.json is shown in the Listing \ref{lst:package.json}, each extension has its own \texttt{package.json}. The extension can define a set of other extensions to be installed along with its installation, and \texttt{extensionPack} \textcolor{red}{\circled{2}} is used to configure that option. \texttt{extensionDependencies} is also used to define other extensions that are required and runtime dependencies to the other extension. Note that there is no consent taken from the user to install extensions that are described in the \texttt{extensionPack} \textcolor{red}{\circled{2}} and \texttt{extensionDependencies} \textcolor{red}{\circled{3}} contexts. The term \texttt{capabilities} is used to define the runtime scope of the extension. Restricted workspace is a VS Code feature\cite{VSCodeRestrictedWorkspace} in which extensions are not treated as a trusted source.

\begin{lstlisting}[caption={This listing shows a portion of a sample package.json file with keys colored in green and values in blue.}, label=lst:package.json]
{
    "main": "./out/extension",
    "scripts": {
        "compile": "tsc -p ./"
    },
    "dependencies": {  := 11
        "@types/vscode": "0.10.x",
        "axios": "0.0.1"
    },
    "extensionPack": [  := 12
        "chrismarti.regex"
    ],
    "extensionDependencies": [  := 13
        "zobo.php"
    ],
    "repository": {
        "url": "https://me.me"
    },
    "capabilities": {
        "untrustedWorkspaces": { := 14
            "supported": "true"
        }
    }
}
\end{lstlisting}

\subsection{Extension Development}

To develop a VS Code extension, developers use a combination of JavaScript or TypeScript and Node.js, leveraging the extensive VS Code API~\cite{VSCodeAPIReference}. The API allows access to various aspects of the editor, including its UI components, workspace data, settings, and more. 
The VS Code APIs, organized into intuitive namespaces, are crucial for extension development. Key among them are \texttt{vscode.workspace}, for workspace functionalities; \texttt{vscode.debug}, for debugging; \texttt{vscode.scm}, for source control management; \texttt{vscode.extensions}, enabling interactions with other installed extensions; and \texttt{vscode.tasks}, for task management and automation within VS Code. 

\subsection{Marketplace}

The Visual Studio Marketplace~\cite{VSCodeExtensionMarketplace} serves as a central hub for developers to discover, download, and publish extensions and other add-ons for a suite of development tools, including Visual Studio Code (VS Code). To ensure the safety and reliability of extensions, the Visual Studio Marketplace scans for viruses on every submitted extension or update~\cite{VSCodeExtensionTrust}. Furthermore, the Marketplace has implemented measures to deter extension authors from name-squatting, safeguarding the integrity of extension names~\cite{VSCodeExtensionTrust}.
Verifying publishers for Visual Studio Code (VS Code) extensions is a fundamental security measure in the Marketplace. This verification process entails adding a TXT record for domain verification, confirming the publisher's authority over their domain~\cite{VSCodeExtensionPublishingVerifyPublisherSection}.
\section{Threat Model}

The extensibility of VS Code allows for a rich development experience but also presents a significant attack surface. Malicious actors can leverage extensions to execute arbitrary code, access sensitive data, or compromise the integrity of the development environment. 
The threat model of the current analysis is guided through the main three questions we mentioned in the beginning:

\begin{packed_enum}
    \item Can an extension be malicious and pose a threat to the host computer? Developer organization?
    \item Can an extension introduce vulnerabilities to the code?
    \item Can an extension leak sensitive information?
\end{packed_enum}

\subsection{Malicious Extensions}

\begin{figure}[!htb]
  \centering
  \includegraphics[width=0.45\textwidth]{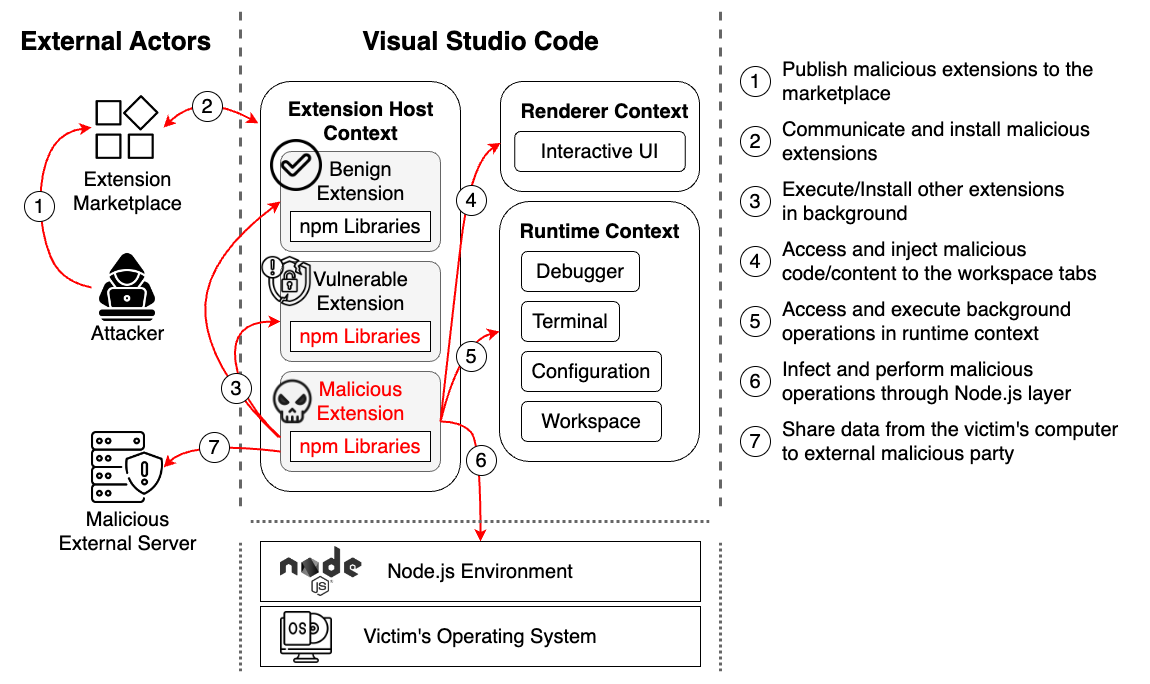}
  \caption{Malicious Extensions}
  \label{fig:vscode_threat_model_malicious}
\end{figure}

In the first scenario (Figure \ref{fig:vscode_threat_model_malicious}), the likely process begins with the attacker publishing a malicious extension to the marketplace. After an unknowing victim installs the malicious extension, the extension can communicate with other installed extensions, including vulnerable extensions, to escalate privileges gain unauthorized access, or directly access underlying unrestricted APIs to carry out the malicious attack. These extensions can then interact with the runtime and renderer contexts, performing malicious operations such as modifying workspace content or running background processes. Crucially, they can leverage Node.js to interact with the victim's operating system and share data with an external server, compromising user privacy and system integrity. These extensions are malicious by nature and designed to harm the host computer, the developer, and/or the developer organization. 

\subsection{Vulnerable Extensions}

\begin{figure}[!htb]
  \centering
  \includegraphics[width=0.45\textwidth]{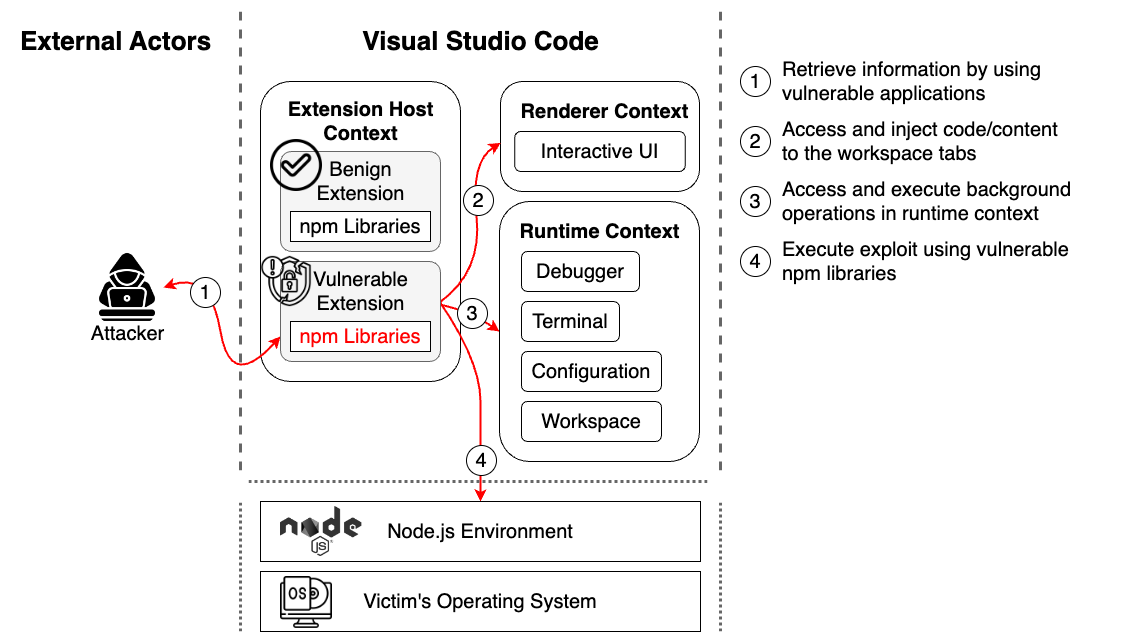}
  \caption{Vulnerable Extensions}
  \label{fig:vscode_threat_model_vulnerable}
\end{figure}

The Vulnerable Extensions scenario (Figure \ref{fig:vscode_threat_model_vulnerable}) illustrates how external attackers can exploit vulnerabilities in legitimate and benign extensions. The attacker may not directly publish a malicious extension but instead uses vulnerabilities in existing extensions to penetrate the developer's host system. These activities include injecting code into workspace tabs and executing background operations, culminating in exploiting npm library vulnerabilities. While the developer of the vulnerable extensions might not be an accomplice to the potential hack, the developer's mistakes are a crucial contributing factor enabling external attackers to compromise the host.

\subsection{Privacy-Invasive Extensions}

\begin{figure}[!htb]
  \centering
  \includegraphics[width=0.45\textwidth]{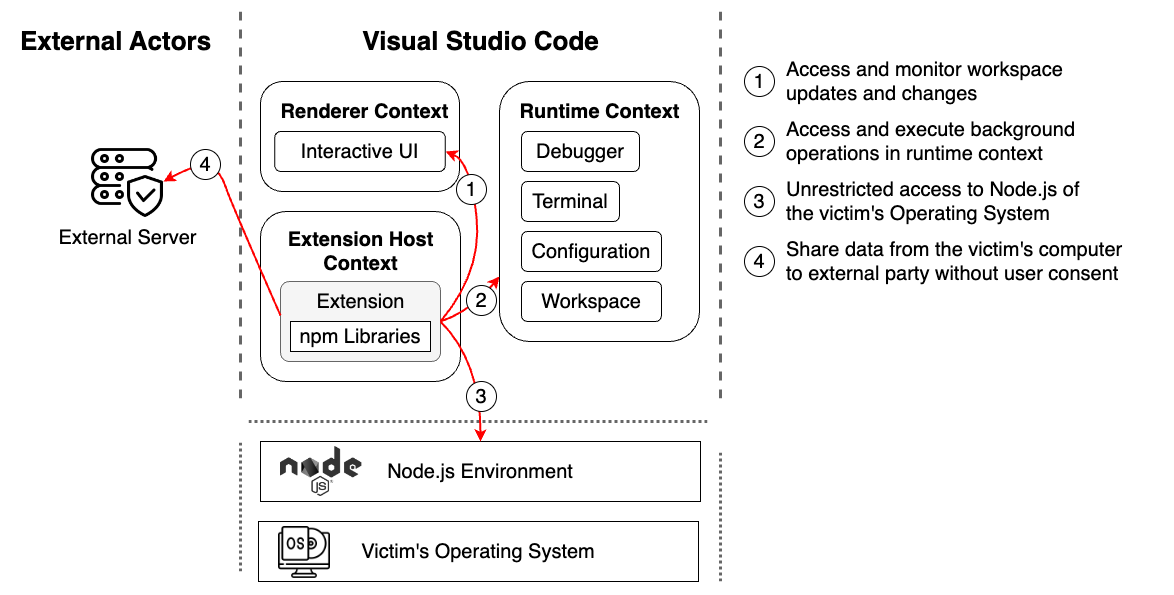}
  \caption{Privacy Invasive Extensions}
  \label{fig:vscode_threat_model_privacy}
\end{figure}

In the Privacy-Invasive Extensions scenario (Figure \ref{fig:vscode_threat_model_privacy}), the threat arises from extensions that overreach their intended scope. They can monitor changes in the workspace, execute unauthorized operations, and transmit data to external servers without user consent, potentially leading to significant privacy breaches. Such breaches carry a significant weight as they could potentially include confidential organizational information. The biggest contributing factors is again the architectural limitations in VS code with a lack of proper developer consent and permissions guarding sensitive information.

\section{Methodology}

Our main objective is to understand the threats posed by VS Code extensions -- we selected the VS Code for two reasons: a) it is the leading IDE/Code Editor used by developers~\cite{StackOverflowSurvey2023}, and we can instrument the Electron-based IDE/Code Editor to understand its execution~\cite{VSCodeRobustExtensibleElectronApp}. Due to the vast array of capabilities available through the platform, we deployed different techniques to filter out extensions and verify the suspicious behavior. We used a mix of static and dynamic analysis. While dynamic analysis exposes the execution behavior, static analysis helps to narrow down the initial questionable list, increasing the scalability of the analysis considerably. We collected a dataset of 52,880 extensions over three months and analyzed them using various tools. This analysis also thoroughly examined the extension code, focusing on the Manifest (\texttt{package.json}) file.

\subsection{Data Collection}

To comprehensively analyze the Visual Studio Marketplace, we developed a crawler to identify and download all extensions listed in the Marketplace. To capture the most up-to-date data, we ran a daily crawler from July 15, 2023, to October 16, 2023, explicitly targeting recently published extensions. After three months of crawling, we ended up downloading 52,880 extensions. To the best of our knowledge, the extension marketplace has no geo-restrictions; hence, we are confident that our initial set of 52,880 is a representative sample of the entire VS Code extension ecosystem.

\subsection{Analysis pipeline: Static}

\begin{figure}[!ht]
  \centering
  \includegraphics[width=0.4\textwidth]{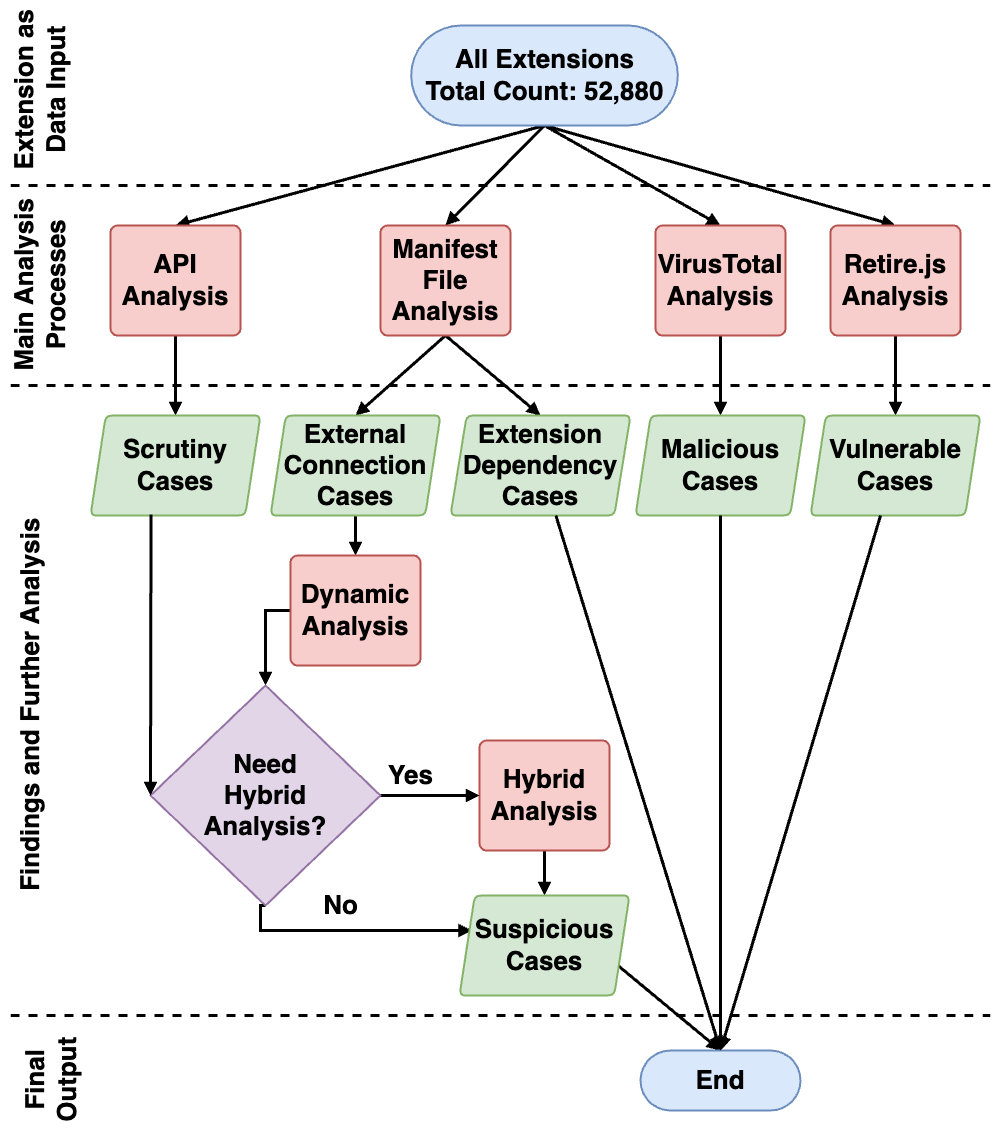}
  \caption{Execution Flow of the Analysis}
  \label{fig:modified_static_analysis_flow}
\end{figure}

In the static analysis phase of our study, we rigorously evaluated all extensions using a combination of tools and code analysis methods: VirusTotal~\cite{VirusTotal}, Retire.js~\cite{RetireJSRepository}, and a combination of VS Code API Usage Analysis and Manifest Analysis. This multi-faceted approach allowed us to capture different threats with a holistic view. Figure \ref{fig:modified_static_analysis_flow} illustrates the static analysis pipeline, delineating the sequential steps and methodologies employed in this phase.

\subsubsection*{VS Code API Usage Analysis}

We analyzed the VS Code API usage of each extension to identify poor security practices or malicious behaviors introduced by the extension developers, potentially raising security threats and risks. 
We then generated an Abstract Syntax Tree (AST) for every JavaScript file within the extracted extension. The AST enabled us to scrutinize each file for VS Code extension API usage and other questionable behaviors. 
\subsubsection*{Manifest Analysis}

We processed the Manifest (\texttt{package.json}) file for all extensions to extract various fields, specifically on \texttt{dependencies}, \texttt{extensionPack}, \texttt{extensionDependencies}, and \texttt{capabilities} for further analysis. 
We aimed to understand how they influence the security posture of each extension and any hidden behavior. We also used manifest analysis to filter any extensions that could share data over the network.

\subsubsection*{VirusTotal Analysis}

VirusTotal~\cite{VirusTotal}, an online platform, facilitates the analysis of files and URLs to identify potential threats such as viruses, worms, trojans, etc. It operates by aggregating a range of antivirus engines and website scanners. We utilized VirusTotal's API to submit all examined extensions for analysis. The generated reports encapsulated the detection outcomes from each participating antivirus engine. 

\subsubsection*{Vulnerability Scan using Retire.js}

Retire.js~\cite{RetireJSRepository} is employed to detect known vulnerabilities, specifically CVEs (Common Vulnerabilities and Exposures) ~\cite{MitreCVE}, in Node.js packages used by VS Code extensions. This tool scans the Node.js dependencies in each extension, revealing their security status. These reports highlight any known vulnerabilities in the Node.js dependencies.
\subsection{Analysis pipeline: Dynamic}

\begin{figure}[!ht]
  \centering
  \includegraphics[width=0.30\textwidth]{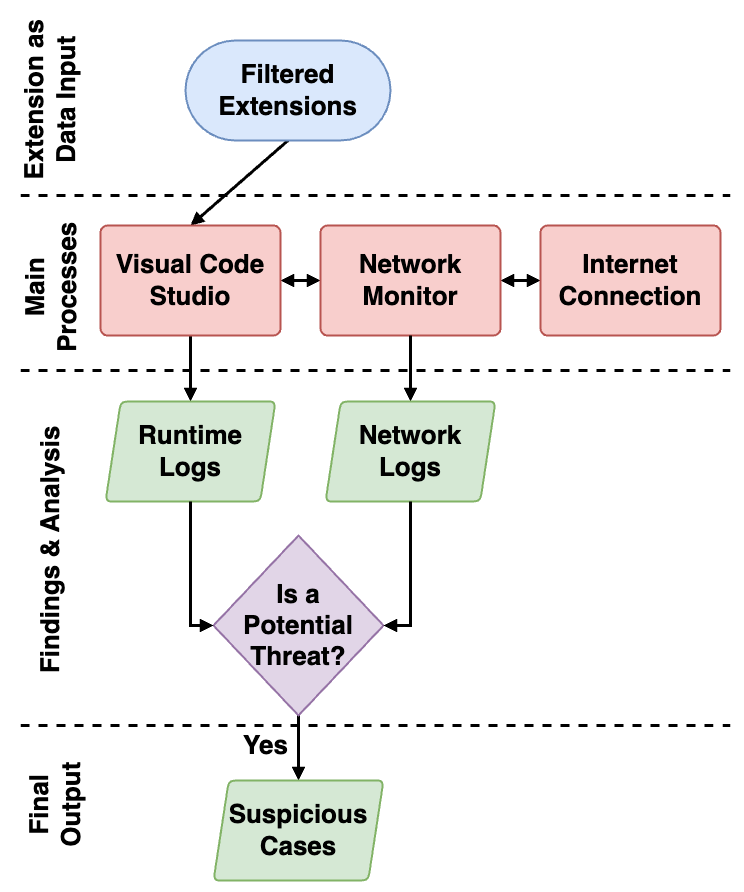}
  \caption{Dynamic Analysis Flow}
  \label{fig:dynamic_analysis_flow}
\end{figure}

We ran the extensions in a customized VS Code editor with network monitoring capabilities using MITM proxy~\cite{Mitmproxy}. Our dynamic analysis pipeline is depicted in Figure \ref{fig:dynamic_analysis_flow}, which provides a detailed illustration of the process. This approach enabled us to capture and analyze the in-situ behavior of the extensions meticulously. 

\subsubsection*{Extensions Selection for Dynamic Analysis}

We selectively focused on extensions that create external connections (network requests), recognizing their greater potential to inflict arbitrary damage on code, developers, host systems, and organizations, as detailed in Section 3.4. Since VS Code does not have a built-in API for establishing external connections, extensions commonly utilize third-party Node.js packages. Based on these findings, we selected 2,365 extensions for in-depth dynamic analysis. Additionally, we chose several other extensions suspected of harboring malicious behaviors, through the static analysis stage. In total, we selected 2,698 extensions for dynamic analysis.

\subsubsection*{VS Code Instrumentation}

We developed an instrumented version of VS Code, utilizing VS Code 1.80~\cite{VSCodeGHRepo} as the base. This instrumented version was designed to capture all VS Code API calls comprehensively. We implemented monitoring mechanisms for terminal access and other resource accesses. These monitoring capabilities were active during both the installation and execution phases of each extension. 

\subsubsection*{Network Request Monitoring}

We utilized a pre-configured mitmproxy proxy~\cite{Mitmproxy} to monitor web traffic generated by the extensions, encompassing protocols like HTTP, WebSockets, and other SSL/TLS-protected protocols. We installed a mitmproxy certificate on the hosting environment to enable the decryption of HTTPS traffic.

\subsection{Analysis pipeline: Summary}

We first ran the static analysis to detect suspicious code or unusual coding patterns. 
These findings were then further scrutinized through dynamic analysis for confirmation. During dynamic analysis, we captured network requests from extensions, including IPs and URLs to which the extensions connect. We then evaluated their maliciousness using threat intelligence resources. Conversely, insights gained from dynamic analysis also informed subsequent static analysis - doing threat analysis using VT for URLs uncovered in the dynamic analysis. 

\section{Visual Studio Code Security Analysis}

\begin{figure}
    \centering
    \includegraphics[width=0.4\textwidth]{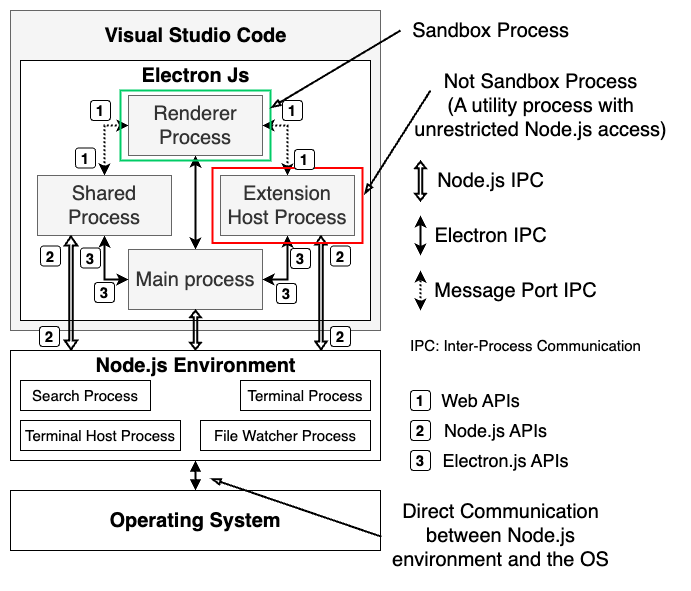}
    \caption{Sandboxed Process Architecture of VS Code.}
    \label{fig:layered_sandbox_process_architecture_vscode}
\end{figure}

This section analyses the security measures in VS Code and their limitations. Figure~\ref{fig:layered_sandbox_process_architecture_vscode} shows the current architecture of the Processes in VS Code. VS Code is based on Electron framework~\cite{ElectronJS}. VS Code architecture, however, has a significant influence from the Chromium browser~\cite{Chromium} and its' V8 JavaScript engine~\cite{V8Dev}. 
However, VS Code extensions are executed on top of a Node.js environment.

\subsection{Sandboxing}

Sandboxing enables untrusted code to provide functionality in a controlled manner containing any harmful behavior of the untrusted code~\cite{AndroidSandboxingUSENIX}. In the current VS Code architecture, only the \textit{Renderer Process} (Figure~\ref{fig:layered_sandbox_process_architecture_vscode}) is sandboxed containing any harmful behavior within the \textit{Rendered Process} which hosts all the VS Code functionalities and the VS Code UI.

The current VS Code architecture does not sandbox \textit{Extension Host Process}, which hosts all the extensions. As per official documentation~\cite{VSCodeSandboxBlog}, providing a scalable solution with full Node.js support is the reason for separating the extension process from the sandboxed \textit{Rendered Process}. The main caveat that comes with that decision is the \textit{Extension Host Process} can access everything from the VS Code data to the host computer files and data. 

As per Electron documentation~\cite{ElectronUtilityProcess}, the \textit{Extension Host Process} hosts a variety of functions, including management of untrusted services, execution of CPU-intensive tasks, and oversight of crash-prone components. This suggests that VS Code extensions, hosted within the \textit{Extension Host Process}, are, rightly so, treated as `untrusted services' -- common wisdom suggests that any third-party code execution should be treated with caution.

\subsubsection*{Untangling the Sandbox}

Security researchers have reported instances where malicious extensions have caused damage in these areas since 2018~\cite{VSCodeIssue52116, goldman2023trust, checkpoint2023vscode}. Letting extensions have a free pass is likely a strategic and marketing decision to promote a robust extension ecosystem that will eventually positively impact the VS Code market share. We also believe that fixing the issues presented in this work is non-trivial which will require a complete re-design of the VS Code architecture, preserving the current functionality of the extensions.

\subsection{Perils of Unchecked Execution}

VS Code adopts a trust-based model in its extension ecosystem, characterized by broad access privileges granted to extensions. These privileges extend to potential interactions with the host system and network, all granted without any notice to the developer. Such sensitive access, while facilitating significant functional enhancement and customization capabilities, simultaneously introduces potential security risks especially if the code is not properly vetted.

This operational paradigm of VS Code stands in stark contrast to the permission-based models employed in platforms like Android and Chrome. In these systems, explicit user permissions are a prerequisite for apps or extensions to access certain functionalities or data.

Upon installation, a VS Code extension typically gains unrestricted access to the VS Code API. This level of access can be perilous if the extension is either malicious or becomes compromised due to its vulnerabilities. Possible threats include arbitrary code execution, unauthorized file manipulation, and data theft or corruption. The unregulated access to APIs raises the possibility of extensions misusing these interfaces, performing actions well beyond their advertised scope. For instance, an extension ostensibly designed for code formatting could hypothetically access and modify unrelated files in the user’s workspace or interact with other extensions in a manner not initially intended -- we have found similar use cases in the current marketplace.

The following list is some of the sensitive API capabilities that are not guarded by any checks or permissions.

\begin{packed_item}
    \item Run potentially sensitive terminal commands.
    \item Modify source code or the commit history without any developer knowledge.
    \item Capture authentication tokens or credentials.
    \item Interfere with other installed extensions.
    \item Modify the behavior of the app during debugging.
\end{packed_item}

The absence of any sort of permission in VS Code keeps developers completely in the dark about extension's execution.

\subsection{Third Party Dependencies}

VS Code extensions, similar to standard Node.js applications, are capable of incorporating third-party Node.js dependencies. While this integration undoubtedly enhances flexibility and promotes the reuse of code, it also introduces a spectrum of security concerns. Literature has documented the escalation of security threats associated with third-party Node.js dependencies (npm)~\cite{Npm_Package_Dependency_Network, npm_ecosystem, dependency_based_nodejs_attacks, zerouali2022impact}.

\subsection{Network Requests}

There is no specific built-in VS Code API for handling network requests. Extensions utilize various third-party Node.js packages to perform network operations. Network operations can be used for potential misuse like data exfiltration, downloading malicious payloads, or establishing command and control communication. \texttt{axios}, \texttt{node-fetch}, \texttt{request}, are Node.js packages commonly used for network requests in VS Code extensions. 

\subsection{Untrusted Workspaces} The VS Code has a restricted mode where all extensions are blocked. For an extension to run on the restricted mode, one must have this property as \textit{true}. If it is \textit{false}, then the extension becomes non-operational in Restricted Mode until Workspace Trust is granted, or \textit{limited}, it will be partially functional in Restricted Mode with certain features disabled. The default value will be \textit{false} if there is no entry. Given that this is a flag the developer voluntarily sets, a malicious developer can set it to true so that the malicious extensions can still be active in the restricted mode of the VS Code. Our testing found 542 extensions with the property set as \textit{true}.

\subsection{Visual Studio Marketplace}

The Visual Studio Marketplace, as the central hub for VS Code extensions, implements security measures to safeguard the integrity and safety of its extensions.

\noindent \paragraph{\textbf{Virus Scanning}}: The marketplace has an extensive virus scanning protocol~\cite{VSCodeExtensionTrust}. Every new submission and update undergoes a thorough virus scan before public release. 
Despite these scans, our analysis found dozens of potentially malicious extensions in the marketplace.

\noindent \paragraph{\textbf{Preventing Name Squatting}}: The Marketplace also prevents ``extension name squatting'' \cite{VSCodeExtensionTrust} to protect users from deceptive practices where unauthorized parties use similar names to popular extensions. This measure is crucial for protecting users from downloading counterfeit extensions.

\noindent \paragraph{\textbf{Publisher Verification}}: This confirms the publisher's authority over their domain, and verified publishers receive a badge in the Marketplace, denoting their verified status. However, we found that 7\% of the tagged extensions for harmful content had the domain verified badge.

\noindent \paragraph{\textbf{Reactive Protocols - Reporting and Kill List}}: In response to security threats, the Marketplace removed such malicious extensions and the implementation of a ``kill list''. This list ensures that compromised extensions are automatically uninstalled from user environments. 

\subsection{Not so Helpful Extension Page}

It's important to consider several key aspects that are often not explicitly detailed on the extension page:

\begin{packed_item}
    \item Many extensions do not provide specific information about their security or privacy practices. Concerned Developers might appreciate the knowledge of who is getting what data about them, their code, or the organization.
    \item It is vital that the developers have full transparency on resource accesses before they make an informed decision to install an extension.
    \item Listing third-party dependencies is also critical to make an informed decision.
\end{packed_item}

It is imperative that developers are given all the possible information to make informed decisions to make sure the development environment is not compromised.

\section{Results}

\begin{table}[]
\centering{
    \caption{Results of the VT Analysis of Extensions }
    \begin{tabular}{|c|c|r|}
    \hline
    \textbf{\begin{tabular}[c]{@{}c@{}}Min VT Positive\\ Engine Count\end{tabular}} &
      \textbf{\begin{tabular}[c]{@{}c@{}}Number of\\ Extensions\end{tabular}} &
      \multicolumn{1}{c|}{\textbf{\begin{tabular}[c]{@{}c@{}}Cumulative\\ Install Count\end{tabular}}} \\ \hline
    1 & 835 & 122,489,729 \\ \hline
    2 & 163 & 1,840,317 \\ \hline
    3 & 62  & 693,147 \\ \hline
    4 & 26  & 385,408 \\ \hline
    \end{tabular}
}
\label{table:vt_analysis}
\end{table}

This section presents the results of our static and dynamic analysis of the 52,880 extensions. We aim to delineate the security risks, threats, practices, and any malicious usage identified in the extensions. These findings focus on aspects that could cause arbitrary damage to code, host systems, developers, or organizations.

\begin{table}[]
\caption{Overall Summary of Results by Threat Model and Suspicious Type.}
\begin{tabular}{|cl|r|r|}
\hline
\multicolumn{1}{|c|}{Threat}                          & \multicolumn{1}{c|}{Suspicious Type}                                              & \multicolumn{1}{c|}{\begin{tabular}[c]{@{}c@{}}Extension\\ Count\end{tabular}} & \multicolumn{1}{c|}{\begin{tabular}[c]{@{}c@{}}Cumulative \\ Install Count\end{tabular}} \\ \hline
\multicolumn{1}{|c|}{\multirow{6}{*}{Malicious}}      & \begin{tabular}[c]{@{}l@{}}Degrading the Security \\ Posture\end{tabular}       & 69                                                                             & 2,809,972                                                                                \\ \cline{2-4} 
\multicolumn{1}{|c|}{}                                & Critical File Access                                                            & 14                                                                             & 1,564,468                                                                                \\ \cline{2-4} 
\multicolumn{1}{|c|}{}                                & VT \textgreater{}= 4 Extensions                                                 & 26                                                                             & 385,408                                                                                  \\ \cline{2-4} 
\multicolumn{1}{|c|}{}                                & \begin{tabular}[c]{@{}l@{}}VT \textgreater{}= 4 Network\\ Requests\end{tabular} & 8                                                                              & 6,239                                                                                    \\ \cline{2-4} 
\multicolumn{1}{|c|}{}                                & Market Missuse                                                                  & 42                                                                             & 254,232                                                                                  \\ \cline{2-4} 
\multicolumn{1}{|c|}{}                                & Concealed Operations                                                            & 18                                                                             & 145,047                                                                                  \\ \hline
\multicolumn{1}{|c|}{Vulnerable}                      & Extensions with CVEs                                                            & 2,620                                                                          & 51,952,070                                                                               \\ \hline
\multicolumn{1}{|l|}{\multirow{3}{*}{API \& Privacy}} & Tracking                                                                        & 49                                                                             & 3,107,508                                                                                \\ \cline{2-4} 
\multicolumn{1}{|l|}{}                                & Code Sharing                                                                    & 108                                                                            & 560,666                                                                                  \\ \cline{2-4} 
\multicolumn{1}{|l|}{}                                & Data Sharing                                                                    & 15                                                                             & 504,835                                                                                  \\ \hline
\multicolumn{2}{|c|}{Total}                                                                                                             & 2969                                                                           & 61,290,445                                                                               \\ \hline
\end{tabular}
\end{table}

\subsection{Malicious Extensions}

\subsubsection*{Extension File}
Our primary discovery is identifying malicious extensions by analyzing VirusTotal (VT) data. Upon scrutinizing the VT results, we initially identified 835 extensions flagged as potentially harmful by at least one antivirus engine. However, for the scope of this research, we have adopted a stringent criterion for classifying an extension as malicious: it must be marked by a minimum of four engines on VirusTotal. This methodology is in line with the best practices recently advocated in the malware research community~\cite{vt_postive}. Table \ref{table:vt_analysis} illustrates the results of our VirusTotal analysis.

Out of the 835 extensions initially flagged, we discovered that 667 had at least one previous version. Our analysis of these versions revealed that 102 extensions had at least one prior malicious version.

\subsubsection*{Network Request}

\begin{table}
\centering
\caption{Extensions communicating with VT count greater than or equal 4 domains}
\begin{tabular}{|l|l|r|} 
\hline
\textbf{Extension Name}                         & \textbf{Domain}      &  \textbf{Install Count}  \\ 
\hline
search-pawn-package & sampctl.com          & 1,977            \\
FridaExtension      & fridaplatform.online & 1,823            \\
OI Wiki             & oi-wiki.org          & 1,723            \\
Hey              & hey.network          & 559             \\
SnippetDrop         & snippetdrop.com      & 106             \\
Get rich overnight  & gateio.ch            & 27              \\
VSQuote             & type.fit             & 21              \\
Code Naming Conventions              & openai-proxy.com     & 3               \\
\hline
\end{tabular}
\label{table:vt_url}
\end{table}

During our dynamic analysis, we captured the network requests made by the extensions. We separated IPs and URLs from the dataset and applied a set of heuristics to filter them for further analysis. We filtered out all the local IP addresses from the IP set, resulting in 72 direct external IPs for further examination. Regarding the URLs, we extracted the domains and filtered them using the Cloudflare Radar~\cite{CloudflareRadar} Top 50,000 Domains. We selected an exclude list from the Radar Top 50,000 Domains for further analysis, which yielded 493 URL domains in the dataset. We then submitted all filtered IPs and URLs to VirusTotal to obtain reports, which were analyzed to check for any malicious behavior associated with the IPs and URLs.

After the VT analysis, we found 8 recipients from 8 extensions with more than 3 VT engines flagging them as malicious (Ref Table~\ref{table:vt_url}). Some of these extensions have a significant number of downloads, risking a sizable population of developers. We found nothing sensitive shared with those domains except for one domain receiving search terms.

\subsection{Vulnerable Extensions}

One of our principal findings stems from identifying vulnerable extensions through the Retire.js vulnerability scanning process. We discovered 54 distinct vulnerabilities, each associated with a CVE, within Node.js packages used by these extensions. The severity levels of these vulnerabilities are categorized as low (1), medium (34), and high (19). We identified 5,775 instances of these CVEs across 2,620 extensions with a cumulative install count of 151,952,070. Further analysis revealed that these CVEs originated from 28 unique Node.js packages. 
All VS Code extensions are installed in the $\sim$\texttt{/.vscode/extensions/} folder. If a malicious developer has installed any of the above-mentioned vulnerable extensions, a malicious extension could access this folder, import vulnerable Node.js packages, and exploit them, potentially causing arbitrary damage to the code, host systems, developers, or organizations. We verified that 19 CVEs with a high severity level are exploitable via a demo VSCode extension that can import vulnerable Node.js packages from vulnerable extensions.
Additionally, the ease with which malicious entities can access and install these vulnerable extensions exacerbates the issue. This lack of proactive alerts or guidance poses a significant risk, potentially exposing codebases, developers, host systems, and organizations to security threats inherent in these extensions.

\subsection{Silent Extension Installation}

\begin{table}
\caption{Results of Silent Extension Installation cases.}
\centering
\begin{tabular}{|l|r|}
\hline
\multicolumn{1}{|c|}{\textbf{\begin{tabular}[c]{@{}c@{}}Suspicious Cases\\ of Silent Installation\end{tabular}}} &
  \multicolumn{1}{c|}{\textbf{\begin{tabular}[c]{@{}c@{}}Number of\\ Extensions\end{tabular}}} \\ \hline
  \begin{tabular}[c]{@{}l@{}}Extensions that are installing other \\ extensions \end{tabular}    & 4,317  \\ \hline
\begin{tabular}[c]{@{}l@{}}Extensions that are installed by other \\extensions \end{tabular}       & 4,618  \\ \hline
Chain extensions (E1 $\to$ E2 $\to$ E3)                 & 2,327  \\ \hline
\begin{tabular}[c]{@{}l@{}}With external connections \\ (for `installed by' cases) \end{tabular}  & 325   \\ \hline
\begin{tabular}[c]{@{}l@{}}VT Positive Cases \\ (for `installed by' cases) \end{tabular}     & 92    \\ \hline
\begin{tabular}[c]{@{}l@{}}CVE Positive Cases \\ (for `installed by' cases) \end{tabular}          & 325   \\ \hline
\end{tabular}
\label{table:ex_dep_cases_analysis}
\end{table}

We have discovered that in VS Code, extensions can install additional extensions through three different methods. These include specifying dependencies in the manifest file using \texttt{extensionPack} or \texttt{extensionDependencies}, and utilizing the VS Code Command API. A notable command within this API is \texttt{vscode.commands.executeCommand}, which can invoke \texttt{workbench.extensions.installExten- sion}. This function allows for the silent installation of extensions without the need for user consent. Our findings indicate that 4,317 extensions engage in this behavior, installing other extensions covertly. This includes the creation of extension chains (E1 $\to$ E2 $\to$ E3) and instances where an extension without external connections installs another with such connections. Crucially, we also examined extensions that silently install malicious or vulnerable extensions. Table \ref{table:ex_dep_cases_analysis} provides a detailed overview of these silent extension installation cases with likely harmful content.

\subsection{Suspicious Extensions}

\subsubsection*{Degrading the Security Posture}

In our VS Code API usage analysis of extensions, we identified several suspicious coding patterns that may pose security risks. With the recent wave of supply chain attacks, much of the attention has shifted towards securing the development environment~\cite{supply_chain_security_30_industry}. We found several cases where extensions have severely degraded the security posture of the host machine, which will negatively impact not only the host's machine but could affect the organizational security posture as well. The only reason for this occurrence is the lack of proper containment -- sandboxing -- in the VS Code for untrusted extensions.

We found that 14 extensions directly access SSH private keys or cloud access tokens via file access paths, posing additional security vulnerabilities. This is the most severe threat posed by malicious extensions. Accessing SSH private keys and cloud access tokens has repercussions far beyond an individual's development machine. These tokens can be used to log into highly sensitive servers in the organizations, polluting production artifacts and/or update channels. Recent supply chain attacks have demonstrated that they have successfully targeted developers to exploit their weakness to access tokens, but this way, it can be done with the near certainty of success, which is dreadful.

Another severe threat was intentionally degrading the TLS setup in the environment \texttt{process. env.NODE\_TLS\_REJECT\_UNAUTHORIZED = 0} This particular code snippet turns off TLS/SSL certificate validation, which exposes the extension to severe security risks, including man-in-the-middle attacks and potential data breaches. Our study found that 60 extensions contained this pattern, compromising security. These extensions have a median install count of 314.5. To corroborate our findings, we conducted a dynamic analysis, further verifying the significant risks posed by this practice. We identified several interesting extensions, one of which is `Black Box'~\cite{nikhilmjeby.black-box}. This extension interacts with the OpenAI API and notably disables SSL/TLS certificate verification(130,913 downloads).

We identified eight extensions that implement local proxy servers to enhance their functionality. These proxies are used primarily because they can circumvent the stringent network security measures established in most corporate environments. Such security measures typically include firewalls, intrusion detection systems, and content filtering policies. A VS Code extension could bypass these security protocols by setting up a local proxy, thereby creating a backdoor. This backdoor allows data to be transferred in and out of the network without undergoing standard scrutiny and security checks.

We also discovered that 329 extensions update the VS Code \texttt{settings.json} file, either through the API (using \texttt{vscode.workspace.getConfiguration("config- ").update("new value"))} or by directly modifying the \texttt{settings.json} file. This practice raises significant security risks and threats. An extension can use this feature to turn off any security feature in the VS code again without any developer's consent.

\subsubsection*{Data Sharing}
Our dynamic analysis of 13 VS Code extensions revealed that a significant amount of data is transmitted over the internet.

VS Code extensions share a variety of information, including device specifics like unique Device ID, language settings, screen dimensions, and timezone. Operating System (OS) details are also communicated such as the username, hostname, OS name, architecture, version, and home directory path. Moreover, details about the working project are transmitted, such as the project name, directory, source code, timestamp, and analytic data which include Google Analytics, workspace dependencies, user actions, GitHub account information, log events, and stack traces.

These kinds of information at the hands of the wrong actors can be devastating for the developers and the organizations. These metadata can reveal a lot of information about the proprietary developments that are yet to be released.

\subsubsection*{Tracking}
Our investigation identified 46 extensions that actively monitor and transmit data regarding user, device, project, session, coding time, user IP, user actions, commit logs, and session details. Alarmingly, 28 of these extensions fail to disclose their tracking practices to developers. The `Code Time'~\cite{swdcVsCode} extension, with 425,386 downloads offers programming metrics, infringes upon privacy by clandestinely collecting information such as usernames, device hostnames, timezones, and detailed project directories alongside project names, all without user awareness.

\subsubsection*{Code Sharing}
In the course of our research, we identified a total of 108 extensions within the VS Code ecosystem that engage in the sharing of source code over the internet. 

We discovered a wide range of code-sharing practices with external services. Notably, 78 extensions actively transmit data to public and private large language models, specifically focusing on two extensions that share code with locally run language model servers. Recently, a few cases appeared to have leaked sensitive code to LLMs due to inaccurate configuration; with an increasing number of extensions now exploiting LLMs, it is worth noting that for almost all of the extensions, developers lack any control over configuring the LLMs that the extension use. However, within this subset, 27 extensions offer limited information about the recipients of this code on the VS Code Marketplace extension's page. We also identified three extensions transmitting selected code to servers for synonym finding and 17 extensions sending code to translation APIs. Concerningly, in both these groups, several extensions (two in the former and 12 in the latter) lack explicit disclosure regarding the destinations of the transmitted code.

Furthermore, three extensions designed for code deployment clearly state the destination of the deployed code. In code snippet sharing, four extensions stand out for their core functionality of sharing code, with one using an encryption scheme for the source code. An extension facilitating interaction with an online code playground and another intended for code vulnerability scanning must adequately disclose their data-sharing practices on their marketplace pages. Finally, an extension focused on variable naming, transmitting selected text without proper validation, is also reticent about the destination of this data. These findings underscore the diverse nature of code-sharing functionalities among VS Code extensions, marked by varying degrees of transparency and leading to potential security and privacy concerns in the ecosystem.

Further, we found extensions transmitting personal information and logs over the internet. Out of 12 such extensions analyzed, 11 shared sensitive data, including stack traces with error logs, GitHub usernames, user emails, and extension logs, without prior disclosure to the developers. This is equally concerning as code sharing since stack traces and execution logs can still expose sensitive information.

\subsubsection*{Suspicious APIs}
Our API analysis revealed that the misuse of certain APIs can cause arbitrary damage to host codebases, developers, host systems, and organizations. We utilized dynamic analysis and runtime logs to verify the extent of API usage in suspicious extensions. Detailed information for these APIs is provided in Table ~\ref{table:api_misuse}, which highlights the potential security risks associated with their misuse.

\begin{table*}[]
\caption{Analysis of API Misuse in Suspicious Extensions. This table details the usage, number of affected extensions, and median installation counts for critical APIs that, when misused, can pose significant threats to system security.}
\begin{tabular}{|l|l|l|l|l|}
\hline
\multicolumn{1}{|c|}{API}                                                                                                      & \multicolumn{1}{c|}{Summary}                                                                                                                                                                                         & \multicolumn{1}{c|}{\begin{tabular}[c]{@{}c@{}}Extension \\ Count\end{tabular}} & \multicolumn{1}{c|}{\begin{tabular}[c]{@{}c@{}}Usage \\ Count\end{tabular}} & \multicolumn{1}{c|}{\begin{tabular}[c]{@{}c@{}}Install Count\\ (Median)\end{tabular}} \\ \hline
workspace.fs                                                                                                                   & \begin{tabular}[c]{@{}l@{}}This file system API can read and write files. If misused, it could lead to \\ unauthorized file access or modification.\end{tabular}                                                     & 469                                                                             & 35,782                                                                      & 648                                                                                   \\ \hline
workspace.createFileSystemWatcher                                                                                              & \begin{tabular}[c]{@{}l@{}}Monitoring file system changes can be exploited if the data is used \\ maliciously.\end{tabular}                                                                                          & 335                                                                             & 7,179                                                                       & 1,235                                                                                 \\ \hline
workspace.applyEdit                                                                                                            & \begin{tabular}[c]{@{}l@{}}This can modify files in the workspace, posing a risk if used to make \\ unauthorized edits.\end{tabular}                                                                                 & 314                                                                             & 7,597                                                                       & 865                                                                                   \\ \hline
workspace.findFiles                                                                                                            & \begin{tabular}[c]{@{}l@{}}This API can access various files within the workspace, which could \\ potentially expose sensitive information.\end{tabular}                                                             & 230                                                                             & 5,947                                                                       & 1041                                                                                  \\ \hline
window.activeTextEditor                                                                                                        & \begin{tabular}[c]{@{}l@{}}Improper use of this API could lead to unauthorized access or \\ manipulation of the active text editor content.\end{tabular}                                                             & 1,136                                                                           & 73,073                                                                      & 287                                                                                   \\ \hline
window.createTerminal                                                                                                          & \begin{tabular}[c]{@{}l@{}}This API allows the creation of a terminal within VS Code with \\ full access to the host machine. A potential risk is executing arbitrary\\  commands through the terminal.\end{tabular} & 267                                                                             & 3,743                                                                       & 491                                                                                   \\ \hline
window.createWebviewPanel                                                                                                      & \begin{tabular}[c]{@{}l@{}}Webviews can display arbitrary content, including potentially\\ malicious HTML or JavaScript.\end{tabular}                                                                                & 681                                                                             & 17,488                                                                      & 346                                                                                   \\ \hline
authentication.getSession                                                                                                      & \begin{tabular}[c]{@{}l@{}}Accessing authentication sessions can pose risks if the session \\ data is mishandled or exposed.\end{tabular}                                                                            & 103                                                                             & 2,978                                                                       & 572                                                                                   \\ \hline
extensions.getExtension                                                                                                        & \begin{tabular}[c]{@{}l@{}}Involves accessing details about installed extensions, which could \\ be misused to target specific extensions or behaviors.\end{tabular}                                                 & 488                                                                             & 11,559                                                                      & 1,252                                                                                 \\ \hline
env.openExternal                                                                                                               & \begin{tabular}[c]{@{}l@{}}This API is used to open URLs or files in the default external \\ application. If not properly validated, it could be exploited to open \\ malicious links or files.\end{tabular}         & 554                                                                             & 17,703                                                                      & 527                                                                                   \\ \hline
env.clipboard                                                                                                                  & \begin{tabular}[c]{@{}l@{}}Access to the clipboard could lead to the exposure of sensitive \\ data copied by the user.\end{tabular}                                                                                  & 301                                                                             & 8,060                                                                       & 804                                                                                   \\ \hline
\begin{tabular}[c]{@{}l@{}}env.sessionId, env.machineId,\\ env.uiKind, env.remoteName,\\ env.appRoot, env.appHost\end{tabular} & \begin{tabular}[c]{@{}l@{}}These environment variables provide system and session information \\ which, if exposed, can be sensitive\end{tabular}                                                                    & 254                                                                             & 9,702                                                                       & 5972                                                                                  \\ \hline
\end{tabular}
\label{table:api_misuse}
\end{table*}

\subsubsection*{Concealed Operations}
In our dynamic analysis, we observed several extensions with concealed activities. One extension was identified as redirecting developers to a malicious webpage, consequently leading to the installation of an unwanted browser plugin. Another extension enabled access to a private extension marketplace within VS Code. Moreover, our investigation revealed 12 extensions clandestinely downloading third-party libraries, command-line tools, and executables without obtaining user consent or providing any notification. In addition, we discovered three extensions initiating terminals within VS Code stealthily by employing the \texttt{hideUser} parameter in the \texttt{createTerminal} API, thus operating unbeknownst to the user. 

\subsubsection*{Network Capabilities}
We discovered the F5 NIM extension~\cite{F5NIMExt} by F5DevCentral, serving as an NGINX Instance Manager with 1502 installs. The Visual Studio, Code extension marketplace, is described as a tool enabling the discovery and management of NGINX instances and configuration files. Upon executing the "Start Scan" command, developers initiate a scan of NGINX instances, launching an underlying network port scan. This was never explicitly mentioned. 

\subsubsection*{Market Misuse}
In our analysis, we encountered several instances indicative of marketplace misuse. We identified 16 extensions published predominantly for testing purposes, as evidenced by their lack of specific functionalities and apparent experimental nature. Additionally, our investigation revealed 69 extensions presenting multiple identical versions, characterized by shared logos, identical descriptions, and only minor variations in titles or publishers, hinting at potential duplication issues. After analyzing the \texttt{repository} field in the \texttt{package.json}, which contains the repository URL, we identified that 9,185 extensions were published without a repository. Of these, 75 extensions were flagged as malicious by VirusTotal (VT), and 142 were associated with known Common Vulnerabilities and Exposures (CVEs). Further analysis of the repository URLs revealed that 120 extensions were clones of other extensions. Furthermore, we observed four extensions that bundled Node modules, EXE, JAR files, and JRE folders, leading to abnormally large extension sizes, a concern we've labeled `oversized extensions.'
\section{Related Work}

There is limited research on the security and privacy of IDEs and code editors, particularly concerning their plugins or extensions. Elizabeth Lin ~\cite{lin2024untrustide}, identified 716 dangerous data flows and 21 verified extension vulnerabilities with proof-of-concept exploits in VS Code extensions. 
The investigation did not cover the marketplace for malicious or suspicious extensions. A. David proposed implementing a permission system for VS Code extensions~\cite{msc_thesis_on_vscode_extensions}. They only evaluated  56 extensions and faced challenges in fully mapping npm package usage to specific permissions. 
Jin et al.~\cite{ndss_Jin0CD0W23_electron_applications} identify security vulnerabilities in the VS Code Markdown editor.

VS Code extensions are based on Node.js, and a few studies have investigated the security risks and vulnerabilities associated with Node.js packages~\cite{Npm_Package_Dependency_Network, npm_ecosystem, dependency_based_nodejs_attacks, zerouali2022impact}. The authors have highlighted a few issues, such as a lack of regular updates, compromised dependency trees, and widespread vulnerabilities in NPM packages. 

Numerous studies have analyzed various third-party integration tools for applications~\cite{sec22_kasturi_wordPress_malicious, zha2022hazard, sec22_shen_yun_persistence_malicious_apps, sec22_koishybayev_GithubCI, CICD_Malware}. Authors have looked into different plugin architectures such as WordPress, Team Chat, and Android Marketplace. 

Ladisa et al. scrutinize OSS supply chains, identifying their complexity as a source of security vulnerabilities~\cite{sok_supply_chains}. Insights from 30 organizations are synthesized, spotlighting challenges like updating vulnerable dependencies, leveraging SBoM for enhanced security~\cite{supply_chain_security_30_industry}. A study on the use of open source components (OSCs) in software projects reveals that their selection often depends on superficial metrics like downloads or GitHub stars disregarding security~\cite{qualitative_open_source_supply_chain}. The significance of Reproducible Builds (R-Bs) in software projects, particularly following the 2020 SolarWinds attack~\cite{solarwind}, is highlighted~\cite{reproducible_builds_for_software_supply_chain_security_conf_oakland_fourne23}.

There is an extensive corpus of literature on the analysis of malicious and vulnerable extensions~\cite{sec17_sanchez_rola_extension_breakdown, mystique, buyukkayhan2016crossfire, hulk_malicious_browser_extensions_sec14_paper_kapravelos, Security_Vulnerabilities_with_VEX, Privacy_Compliance_Violations_among_Browser_Extensions, Fighting_Malicious_Extensions, spying_browser_extensions, browsing_activity, browser_extension_download_patterns}. Studies such as \cite{malicious_browser} and~\cite{EmPoWeb} utilize static analysis to identify extensions that potentially abuse privileges, including improper access to APIs and sensitive user data. Similarly, research like~\cite{Malicious_Browser_Extensions_at_Scale},~\cite{Verified_Security_for_Browser_Extensions}, and~\cite{Privacy_Compliance_Violations_among_Browser_Extensions} use static analysis to detect misuses of permissions.

\section{Discussion}

We have analyzed 52,880 extensions and found 2969 (5.61\%) extensions to be potentially harmful to the developer under five different categories. These extensions have a cumulative install count of 613 Million installs, exposing a significant number of developers worldwide. VS Code being the most popular IDE/Code Editor in the world~\cite{StackOverflowSurvey2023}, the ~6\% likely malicious extensions from the VS Code marketplace is a significant number. This work presents the first holistic analysis of the VS Code extension ecosystem.

Our analysis uncovered issues along two main axes: security of the host, the network, and the organization, as well as problems of privacy and IP of the developer and the organization. Implications of a compromised extension with an ability to steal pretty much anything out of the host machine are dangerous and require a thorough examination to fix it before a real-world attack occurs. Supply chain attack on SolarWinds~\cite{solarwind}, key exposure in Microsoft~\cite{Satter2023MicrosoftHack} exposes the ugly truth of being complacent on developer security.

With the recent high-profile wave of IP theft and supply chain attacks, having untrusted extension code roaming in the host with unparalleled access should be a nightmare for organizations and developers. These unintended ID leaks can lead to sophisticated spear phishing or, even worse, targeted attacks on the host to steal code or other sensitive materials. With extensions having high severity vulnerable code executing with unchecked capabilities, these extensions themselves could be weaponized for IP theft, geopolitical advantages, or criminal gangs. The first and foremost step would be to educate developers and get the relevant stakeholders to act upon it.

Security and privacy literature has been focusing on consumer and organizational security and privacy. It is high time that we also devote our focus to developer security. Literature has looked into secure CI/CD~\cite{sec22_koishybayev_GithubCI, CICD_Malware} and secure update channels~\cite{supply_chain_security_30_industry}, but threat vectors such as developer extensions are usually overlooked. But as the analysis suggests, developer tools and extensions present an equally important threat vector to be concerned with. The current analysis opens up two avenues for further examination: a secure extension/tool architecture with visibility and the ability to audit.

Microsoft, the VS Code team, knew about the potential dangers of the current extension architecture since 2018~\cite{VSCodeIssue52116}. The online documentation suggests their design choices were largely based on creating a capable extension ecosystem with minimal restrictions~\cite{VSCodeSandboxBlog}. This is a sensible decision for creating a vibrant ecosystem. Still, security has to be a critical factor in designing the architecture. It raises the question of the inaction by Microsoft for 5 years, potentially underestimating the grave ramifications.

Moving forward, the solution must be multi-faceted: how can VS Code create a highly capable extension architecture with correct checks and guards following defense-in-depth and least-privilege principles, and how can the developer make informed decisions before installing extensions? The solution should guard different capabilities with permission-guarded APIs so that the developer is also in the loop during the execution. VS Code marketplace should also have a strict structure (such as in Google Playstore) with permission and capability information so that developers can comprehend the extension before deciding which tool to pick. 

This work also calls for more Developer studies on understanding their privacy and security expectations while using developer tools. There can be a lot of lessons learned from Android permission literature~\cite{felt2012android, felt2012android_How_to_Ask_for_Permission, Android_Permissions_Demystified} on understanding different contexts and approaches to effectively seek permission before granting access to sensitive information. Developers are likely to have an opinion on when they want to be prompted for permissions, under what contexts, for what types of capabilities, etc. Developers are also likely to want to have an audit mechanism where they can go back and audit extension history and change their future behavior at a finer level. All of these are major changes that will considerably affect the extension ecosystem. Thus, other factors such as adoption, usability, and retention will come into play in any decision around revamping the current extension architecture.

In conclusion, our analysis reveals that developer extensions pose a credible threat to the developer, the code, the host computer, and the organization. The analysis uncovers five types of questionable behavior: (a) developers have minimal visibility into what extensions are accessing behind VS Code, (b) extensions have unchecked access to the code and the host computer, (c) some extensions purposely degrade the host computers security posture opening up to the possibility of getting hacked, (d) extensions are sharing sensitive developer information, code, execution logs, and stack traces over the internet, and (e) some extensions are either very likely malicious or riddle with vulnerabilities. The work analyzed 52,000+ VS Code extensions, and ~5.6\% of them have suspicious behavior potentially putting over 500 million developers at risk.

\section*{Acknowledgments}
This work is supported by the Australian Research Council Discovery Project (DP210102761).

\bibliographystyle{IEEEtran}
\bibliography{References}

\end{document}